\documentstyle[aps]{revtex}
\begin{document}
\title{Flavor independence and the dual superconducting model of QCD}
\draft
\author{Lewis P. Fulcher}
\address{Physics and Astronomy, Bowling Green State University\\ Bowling
Green, Ohio 43403\\ email: fulcher@newton.bgsu.edu}
\date{\today}
\maketitle
\begin{abstract}
Baker, Ball and Zachariasen have developed an elegant formulation of the
dual superconducting model of quantum chromodynamics (QCD), which allows
one to use the field equations to eliminate the gluon and Higgs degrees
of freedom and thus to express the interaction between quarks as an effective 
potential.  Carrying out an expansion in inverse powers of the constituent
quark masses, these authors succeeded in identifying the central part, the 
spin-dependent part and the leading relativistic corrections to the central 
potential. The potential offers a good account of the energies and
splittings of charmonium and the upsilon system. Since all of the flavor 
dependence of the interaction is presumed to enter through the constituent 
masses, it is possible to test the potential in other systems. Logical 
candidates are the heavy B-flavor charmed system  and the heavy-light systems,  which should be more sensitive to the relativistic corrections. Lattice
gauge calculations furnish an additional point of contact for the components
of the BBZ potential.
Some preliminary calculations of the energies of B and
D mesons are presented and the challenge of agreement with experiment is
discussed. The spinless Salpeter equation is used to account for the
effects of relativistic kinematics.
\end{abstract}
\pacs{12.39.Ki,12.39.Pn,12.38.Aw,12.40.Yx}

\section*{introduction}
\label{sec:intro}
After an extensive investigation Baker, Ball and Zachariasen \cite{ba88,ba91r}
(BBZ) 
have developed a number of arguments to support the use of the dual color 
fields (derived from the color electric vector potential $C_{\mu }$),
instead of the usual Yang-Mills fields (derived from the magnetic vector
potential $A_{\mu }$), to formulate an effective theory of long-distance 
quantum chromodynamics (QCD). The Dyson-Schwinger equation may be used     
to justify some of their arguments, since it shows that corrections to the
leading terms in the $A_{\mu }$ propagator $\Delta _{A}$ are more
singular in the infrared region that the leading term. This behavior
is not true for the $C_{\mu }$ propagator, and thus one should use the dual
fields in the long-distance region. Furthermore, BBZ argue that the connections
between the dual fields and the Yang-Mills fields are nonlocal and require
the introduction of an additional field. They use this requirement to good
advantage by introducing Higgs fields, which allow a symmetry breaking
mechanism to operate and thus confinement is explained as a Meissner effect,
in the spirit of the 't Hooft-Mandelstam conjecture \cite{ho76,ma76},           where a magnetic condensate exerts pressure on the gluon field lines. 
The BBZ Lagrangian leads to an
effective field theory with many of the properties expected from QCD, such
as unitarity, renormalizability and non-Abelian gauge invariance. One
can also show that the theory undergoes a deconfinement transition 
\cite{baprl88}. A recent examination of the foundations of the BBZ model
\cite{ba97} led to a reformulation of the theoretical underpinnings in a
Wilson loop context.

In their more recent work BBZ \cite{ba91,ba93,ba95} have constructed an
interaction potential between a quark source and its antiquark by 
developing an expansion in inverse powers of the constituent masses.
They use the field equations to eliminate the gluon and the Higgs fields and 
express the interaction energy in terms of the quark and antiquark variables.
They can readily identify the central potential, the spin-dependent potentials
and the leading relativistic corrections to the central potential. In this paper
we explore some of the problems that arise when one attempts to use this 
potential in systems where relativistic corrections are more important.
In particular, we attempt to give a
simultaneous account of the properties of charmonium, the upsilon system and
the heavy light systems using the spinless Salpeter equation \cite{ols86}       for relativistic
kinematics and a single set of potential parameters \cite{fu95,fu97}.    
In addressing this issue we will be led to examine questions about the
relationships between the constants in the BBZ potential components and the
parameters associated with mass renormalization \cite{bbp}. We note in
passing the interesting work of Crater and Alstine \cite{cra92,cra94} that
explores the question of a universal quark-antiquark potential in the context
of the Todorov equation and constraint dynamics.

Since spin dependence is not a natural companion to work based on the
spinless Salpeter equation, we will focus on spin-averaged energies as
the point of contact with experiment. This goal is consistent with the
goal of the BBZ investigations to produce an effective theory of long-range
QCD, since spin-dependent splittings are presumably more sensitive to the
short-range behavior of the potentials, where running coupling constant
effects are important.  Although it is possible to accomodate such effects
within the framework of the BBZ model, this involves additional assumptions
\cite{ballcom}.  We have plans to pursue the role of the spin dependence in the
context of the work of Hardekopf and Sucher \cite{su84}.

Our method of solution of the spinless Salpeter equation is based on a
recent improvement \cite{fu93,fu94} of the Rayleigh-Ritz-Galerkin method.    
The purpose of our calculation is to see how successful one can be with the
leading relativistic corrections and what role experiments might play in
constraining these corrections and their relativistic generalizations.
We establish contact with recent precision
lattice gauge calculations of Bali, Schilling and Wachter \cite{bal96,bal97} in
order to have an additional guide for evaluating the relativistic
corrections.

The highpoints of the BBZ formalism are summarized in Sec.\ 
\ref{sec:model}, and the solution of the spinless                       
Salpeter equation is presented in Sec.\ \ref{sec:sol}. Some of the
problems associated with the Salpeter wave functions and the salient
features of the interpretation of the momentum operators are discussed in
Sec.\ \ref{sec:veldepcor}.
Our results and conclusions are presented in Sec.\ \ref{sec:res}.

\section{BBZ model}
\label{sec:model}
The fundamental degrees of freedom of the BBZ model are an octet of dual
gluon potentials $C_{\mu }^{a}\, (a=1,\ldots ,8)$ and a triple of scalar
Higgs fields $\vec B^{a}$. Because of the symmetry breaking mechanism, these
Higgs fields have nonzero vacuum expectation values, which give rise to
effective masses for the dual gluons and two of the scalar Higgs fields.
These effective masses are responsible for the exponential damping of all
fields outside of the region of the sources and the flux tube that connects
them. The model is based on the Lagrangian density,

\begin{equation}
\label{lagden}
{\cal L } = TR \left[ \left( \bf H^{\rm 2} - D^{\rm 2 } \right) + \left(
{\sl D_{\mu}} \vec B \right)^{2} \right] - W(\vec B) ,
\end{equation}
where all the field  variables in Eq.\ (\ref{lagden}) are scalar products with
the SU(3) color matrices $\lambda_{a}$ (e.g. $C_{\mu } \equiv \frac{1}{2}\,     
\Sigma \, C_{\mu }^{a}\lambda _{a}$) and $W(\vec B)$ denotes the Higgs 
potential. The covariant derivative in Eq.\ (\ref{lagden}),
\begin{equation}
\label{covar}
D_{\mu }\vec B \equiv \partial_{\mu }\vec B - ig \left[ C_{\mu}, \vec B \right],
\end{equation}
is necessary for non-Abelian gauge invariance. The electric displacement
{\bf D} and the magnetic field {\bf H} are related to the vector potential
$C_{\mu }$ by the non-Abelian relations,
\begin{mathletters}
\label{fielddef}
\begin{equation}
\bf D = -\nabla \times C - {\rm \frac{ig}{2}} \left[ C, \times C \right],
\end{equation}
\begin{equation}
{\bf H} = -\nabla C_{0} - \partial_{0} {\bf C} - ig \left[ {\bf C}, C_{0}
\right].
\end{equation}
\end{mathletters}
 
To derive the simplest set of field equations with non-Abelian couplings,
BBZ choose a gauge where the components of the Higgs triple point in three
independent color directions, namely, $ \vec B = (\lambda _{7} B_{1},
-\lambda _{5} B_{2}, \lambda _{2} B_{3})$ and all the dual gluon field variables
are proportional to the color matrix $Y = \lambda _{8} / \sqrt{3}$. This
particular choice leads \cite{bapr} to finite results for the commutators
of the first two components of $\vec B$ in Eq.\ (\ref{covar}), but the remaining commutators in the equations above vanish.
Thus all the relationships
in the gluon sector are Abelian. Evaluating the trace of Eq.\ (\ref{lagden})
yields
\begin{equation}
\label{finalden}
{\cal L } = \frac{2}{3} \left( \bf H^{\rm 2} - \bf D^{\rm 2} \right) + 
2\; \partial _{\mu } \vec B \cdot \partial ^{\mu }\vec B +
2\: g^{2} C_{\mu }C^{\mu} \left( B_{1}^{2} + B_{2}^{2} \right) + W(\vec B).
\end{equation}
Since this Lagrangian density is symmetric in the components $B_{1}$ and
$B_{2}$, we may choose them to be equal. The fourth-order term in Eq.\
(\ref{finalden}) leads to effective mass terms in the field equations of
the dual vector potential $C_{\mu }$ and the first two components of $\vec B$,
thus restricting these fields to the flux tube region and its vicinity.

To complete the calculation of the interaction energy of a quark and its 
antiquark, one must introduce these particles as classical sources with
masses $m_{1}$ and $m_{2}$ and spins $\bf \sigma _{\rm 1}$ and 
$\bf \sigma _{\rm 2}$, which are located at $\bf x_{\rm 1}$ and
$\bf x_{\rm 2}$ and connected by a Dirac string. It is important to
isolate the string contributions to the electric displacement and
the magnetic field, $\bf D_{\rm S}$ and 
$\bf H_{\rm S}$, which are produced by the sources,

\begin{mathletters}
\begin{equation}
\rho ({\bf x}) = e \left[ \delta ^{3}({\bf x} - {\bf x_{\rm 1}}) -
\delta ^{3} ({\bf x} -{\bf x_{\rm 2}} ) \right] Y,
\end{equation}
\begin{equation}
{\bf j}({\bf x}) = e \left[ {\bf v_{\rm 1}} \delta ^{3}({\bf x} -
{\bf x_{\rm 1}}) - {\bf v_{\rm 2}} \delta ^{3}( {\bf x} - {\bf x_{\rm 2}})
\right] Y,
\end{equation}
\end{mathletters}
where the velocities $\bf v_{\rm i}$ are simply the time derivatives of the 
quark coordinates $\bf x_{\rm i}$. The relationships 
between the fields and the potentials also include explicit contributions
from the color magnetization {\bf M} and the color polarization
{\bf M}, which arise from the spin dependence of the sources. Thus,
\begin{mathletters}
\label{str}
\begin{equation}
\bf D = -\nabla \times C + D_{\rm S} - P,
\end{equation}
\begin{equation}
\bf H = - \nabla {\rm C_{0}} - \partial_{\rm 0} C + H_{\rm S } + M .
\end{equation}
\end{mathletters}

From the Lagrangian density of Eq.\ (\ref{finalden}) one can derive the field 
equations for the vector potential $C_{\mu }$ and the Higgs fields
$\vec B$. Their couplings to the sources are given by Eqs.\ (\ref {str}), and
thus the energy content of the fields surrounding the sources is
completely determined by the solution to these field equations. 
Substituting these solutions into the expression for the Lagrangian density and
integrating over all space gives the energy of the quark-antiquark
system as a function of the source parameters $\bf r = x_{\rm 1} - x_{\rm 2}$,
$\bf v_{\rm i}$ and $\bf \sigma _{\rm i}$. Thus the potential is given by
\begin{equation}
V({\bf r,v_{\rm i}, \sigma _{\rm i}}) = - \int d^{3}x \: {\cal L} ({\bf
x,v_{\rm{i}},\sigma _{\rm i },C, \partial_{\rm j} C, {\rm \vec B}} ).
\end{equation}

It is straightforward to identify the central potential, the spin-dependent
potential and the velocity-dependent corrections to the central potential,
that is,
\begin{equation}
\label{potent}
V = V_{0} + V_{SD} + V_{v^{2}},
\end{equation}
where the central potential is given by
\begin{equation}
\label{centpart}
V_{0} = A r - \frac {4 \alpha_{S}}{3 r} e^{-0.511\sqrt{A/\alpha_{S}} \; r}
-0.646 \sqrt{A\alpha_{S}}.
\end{equation}
It is not necessary to list the spin-dependent potential here, since it will not
contribute to the spin-averaged energies in the lowest order.
The velocity-dependent corrections to the central potential are given by the
sum of an angular-momentum term and a radial-momentum term, namely,
$V_{v^{2}} = V_{ang} + V_{rad}$. Each of these terms consists of two parts,
\begin{mathletters}
\label{angrad}
\begin{equation}
V_{ang} = V_{+}\:\left[\bf r \times \left(v_{\rm 1} - v_{\rm 2} \right)
\right]^{\rm 2} / 4 \rm r^{2} \; + \; V_{-}\:\left[\bf r \times \left(
v_{\rm 1} + v_{\rm 2} \right)\right]^{\rm 2} / 4 \rm r^{2},
\end{equation}
\begin{equation}
V_{rad} = V_{\|}\:\left[\bf r \cdot \left(v_{\rm 1 } + v_{\rm 2} \right)
\right]^{\rm 2} / 4 \rm r^{2} \; + \; V_{L} \:\left[\bf r \cdot \left(v_{\rm 1}
 - v_{\rm 2} \right) \right]^{\rm 2} / 4 \rm r^{2}.
\end{equation}
\end{mathletters}
The potentials listed in Eqs.\ (\ref{angrad}) are not all independent.
Because of the Lorentz covariance of the theory \cite{ba95,bbp}, two of them
may be expressed in terms of the central potential as follows: 
\begin{equation}
V_{-} = - \frac{1}{2} V_{0} \hspace{.4in}  ;                                   
\hspace{.4in} V_{\|} = V_{-} + \frac{r}{2} \frac{ \partial
V_{0}}{ \partial r},
\end{equation}
BBZ \cite{ba95} list explicit results for the other two,
\begin{mathletters}
\label{rempot}
\begin{equation}
V_{+} = - \frac{2 \alpha_{S}}{3 r} e^{- 1.14 \sqrt{A / \alpha_{S}} \; r}
      -0.208 A r + 1.12 \sqrt{ A \alpha_{S}},
\end{equation}
\begin{equation}
V_{L} = - \frac{ 4 \alpha_{S}}{3 r} e^{- 0.685 \sqrt{A / \alpha_{S}} \; r}
      +0.0885 \sqrt{A \alpha_{S}}.
\end{equation}
\end{mathletters}
Thus only two parameters, the string constant $A$ and the strong coupling
constant $\alpha_{S}$ are required to determine the potential. Of course,
in the center-of-momentum frame the constituent masses become part of the
parametrization through the relations, $\bf v_{\rm 1} = p / \rm m_{1}$
and $\bf v_{\rm 2} = p / \rm m_{2}$. Since agreement with experiment will
not be easy to achieve, it is important to explore the sensitivity of the
calculation to various parts of $V_{v^{2}}$. Thus we report results
below where the form of the longitudinal component is changed to
\begin{equation}
\label{secondform}
V'_{L} = - \frac{ 4 \alpha_{S}}{3 r}.
\end{equation}

and to
\begin{equation}
\label{thirdform}
V''_{L} = - \frac{ 4 \alpha_{S}}{3 r} e^{- 0.685 \sqrt{A / \alpha_{S}} \; r}.
\end{equation}

In the limit of perturbative QCD ($ A \rightarrow 0$) the potentials
$V_{+}$, $V_{-}$, $V_{\|}$, $V_{L}$ all reduce to multiples of the
Coulomb potential. Thus, the central potential and its leading relativistic
correction reduce to the Darwin limit \cite{jack},
\begin{equation}
\label{dar}
V_{0} + V_{v^{2}} \rightarrow - \frac { 4 \alpha_{S}}{3 r} \left[
1 - \frac{1}{2} \; \left( \bf v_{\rm 1} \cdot v_{\rm 2} +
v_{\rm 1} \cdot \hat{r} \; v_{\rm 2} \cdot \hat{r} \right) \right].
\end{equation}

We determine our parameters in the heavy quark systems, where the 
expansion in powers of the inverse masses is expected to have the 
most validity. Using the spin-averaged 1S, 2S and 1P energies of the 
upsilon system \cite{spindep} and the 1S energy of charmonium, we
find that
\begin{equation}
\label{paramet}
A = 0.235 \: GeV^{2}; \; \frac{4}{3} \alpha_{S} = 0.398; \; m_{b} = 
4.710 \: GeV; \; m_{c} = 1.320 \: GeV,
\end{equation}
for the case that $V_{L}$ is given by Eq. (\ref{secondform}) above.
 
\section{Solution of the spinless Salpeter equation}
\label{sec:sol}
Our semirelativistic potential model is based on the Salpeter equation,
$H \Psi_{n} = E_{n} \Psi_{n}$, where
\begin{equation}
\label{salop}
H = \sqrt{m_{1}^{2} + p^{2}} \; + \sqrt{m_{2}^{2} + p^{2}} \;
   + V({\bf r,p }),
\end{equation}
and the potential is given to O($v^{2}$) by Eq.\ (\ref{potent}).
This model may be thought of as a three-dimensional reduction of the
Bethe-Salpeter equation \cite{ols86,lsg}. We will find the exact solution 
to this equation with the central potential of Eq. (\ref{centpart}) using
the approach of Refs.\cite{fu93,fu94} to the Rayleigh-Ritz-Galerkin method, 
where the wave function $\Psi_{n} $ is expanded in terms of a complete set of
basis functions. Then we will add the effects of $V_{v^{2}}$ with a 
first-order perturbation improvement.

The basis functions \cite{ols86,fu93,fu94,kp} used for the calculation          include a spherical harmonic
factor $Y_{\ell m}({\bf \hat{r}})$ and the radial functions,
\begin{equation}
\label{basisfn}
R_{n\ell} = N_{n\ell } \beta^{3/2} (2 \beta r)^{\ell } e^{-\beta r}
            L^{2 \ell + 2}_{n} (2 \beta r),
\end{equation}
where $N_{n \ell }^{2} = 8 (n !)/ \Gamma(n+2\ell +2)$, $\beta$ is a scale
parameter, and the oscillations necessary for completeness are supplied
by the associated Laguerre polynomials $L_{n}^{2\ell +2} (2 \beta r)$. Two
important advantages accompany the use of these basis functions. The first
of these is that all of the basic matrix elements that we require to find the
eigenvalues and the eigenfunctions of Eq.\ (\ref{salop}) can be generated from
straightforward analytic expressions. The second is that the confinement
factors of Eq.\ (\ref{basisfn}) are a sufficienly good approximation to
the actual wave function that the sizes of the requisite matrices are
relatively modest ($20 \times 20$ or $40 \times 40$). Analytic expressions
for the matrix elements of the operators $(p^{2})_{n n'}$, the linear 
potential, $(Ar)_{n n'}$, and the Coulomb potential $(- 4 \alpha_{S} /3r)
_{n n'}$ are listed in Ref.\cite{fu94}. There it is also explained
how one can calculate the matrix elements of the nonlocal kinetic
energy operators in Eq.\ (\ref{salop}) by effectively taking the square
roots of the matrix representations of the operators $m_{i}^{2} + p^{2}$. 

Thus the remaining challenge in finding
the solutions to the Hamiltonian operator of Eq.\ (\ref{salop}) resides in
the Yukawa factor present in the central potential of Eq.\ (\ref{centpart}).
We write the central potential as a Cornell Potential (linear + Coulomb) and
a residual part by adding and subtracting a Coulomb piece, that is,
\begin{equation}
\label{separ}
V_{0} = V_{Cornell} + \frac{4 \alpha_{S}}{3 r} \left( 1 - e^{-\eta r}\right),
\end{equation}
where $\eta = 0.511\sqrt{A/\alpha_{S}}$. To determine the matrix elements of the
residual part, we use the following algorithm:
\begin{enumerate}
\item Generate the matrix elements of the $r_{op}$, that is, the $(r)_{nn'}$.
\item Diagonalize this matrix to find the eigenvalues $r_{k}$, which give a
unique signature of $r_{op}$ in the chosen Hilbert space.
\item From the diagonal elements form the quantities $f(r_{k}) = \left( 1 -
      e^{-\eta r_{k}}\right) / r_{k}$, which constitute a signature of the
	residual part of the potential of Eq.\ (\ref{separ}).
\item Restore to the original basis with the unitary transformation that
	connects the original matrix representative of $r_{op}$ to its
	diagonal signature.
\item Examine the procedure for stability by varying the sizes of the matrices.
\end{enumerate}

In order to have a definitive test for the stability of this five-step
algorithm, it is convenient to have analytic results for some of the 
low-lying S-state matrix elements. With the basis functions of Eq.\             (\ref{basisfn})
it is straightforward to evaluate a few matrix elements of the residual
operator $f(r)$ when $\ell =0$. We list five of these from the first two rows
of the matrix,
\begin{mathletters}
\label{matrixf}
\begin{equation}
(f)_{00} = \frac{\beta \delta (2 + \delta )}{(1 + \delta )^{2}} \: ; \:
(f)_{01} = \frac{\beta \delta^{2} (3 + \delta )}{\sqrt{3} (1 + \delta )^{3}},
\end{equation}
\begin{equation}
(f)_{02} = \frac{\beta \delta^{3} (4 + \delta )}{\sqrt{6} (1 + \delta )^{4}},
\end{equation}
\begin{equation}
(f)_{11} = \frac{\beta \delta (2 + 3 \delta + 4 \delta^{2} + \delta^{3} )}      {(1 +\delta )^{4}}, 
\end{equation}
\begin{equation}
(f)_{12} = \frac{\beta \delta^{2} (4 + 4 \delta + 5 \delta^{2} + \delta^{3} )}   {\sqrt{2}(1 + \delta )^{5}},
\end{equation} 
\end{mathletters}
where $\delta = \eta / (2 \beta ).$ We have verified that the matrix elements
of Eq.\ (\ref{matrixf}) have the correct small $\eta $ behavior, where a
power series expansion is valid, and the correct large $\eta $ behavior,
where the matrix elements must reduce to those of the Coulomb potential.

The results for the stability study of the matrix elements of the function 
$f(r)$ are presented in Table \ref{potmat}. There results are listed for    
five different matrix sizes used to find the eigenvalues of 
$r_{op}$, as discussed in step 2 of the algorithm below Eq.\
(\ref{separ}). Since there is no discernible difference in the matrix elements
listed in the middle four columns and these results agree with the analytic
results, one concludes that using a $40 \times 40$ matrix should generate
very accurate results for $(f)_{nn'}$. The energy eigenvalues listed in
Table \ref{potmat} were obtained with kinetic energy matrices the same size
as the potential energy matrices. For the three smaller cases the Hamiltonian
matrices were also chosen the same size as the kinetic and potential energy 
matrices, but for the two larger cases a $20 \times 20$ matrix was used for
the Hamiltonian matrix. All of these results are consistent with our earlier
study \cite{fu94} of the stability of the algorithm used to take the
square roots of the kinetic energy operator and the choice of $20 \times
20$ as adequate for the Hamiltonian matrix.

In summary our calculation of the energy eigenvalues and eigenfunctions
requires four separate matrix diagonalizations, two to determine the kinetic 
energy operators, one to determine the residual operator $f(r)$, which
represents the difference between the Yukawa potential and the Coulomb
potential, and a fourth to diagonalize the Hamiltonian matrix. For the first
three diagonalizations we use $40 \times 40$ matrices, and for the last we use
a $20 \times 20$ matrix. One can generate accurate results for a range of
scale values. The choice $\beta = 2.0 \:{\rm GeV}$ is a good one for the upsilon
system, but $\beta = 1.0 \:{\rm GeV}$ is better for the heavy-light systems.

\section{Velocity-Dependent Corrections}
\label{sec:veldepcor}
Working in the center-of-momentum frame, where $\bf v_{\rm 1} = p/{\rm m_{1}}$
and $\bf v_{\rm 2} = - p/{\rm m_{2}}$, allows us to write the velocity-dependent
potentials of Eq.\ (\ref{angrad}) in the form,
\begin{mathletters}
\label{com}
\begin{equation}
V_{ang} = \left[ \left(\frac{1}{m_{1}} + \frac{1}{m_{2}}\right)^{2}
\frac{V_{+}}{4r^{2}} + \left(\frac{1}{m_{1}} - \frac{1}{m_{2}}\right)^{2}       
\frac{V_{-}}{4r^{2}}\right]\bf (r \times p) \cdot (r \times p),
\end{equation}
\begin{equation}
V_{rad} = \left[ \left(\frac{1}{m_{1}} - \frac{1}{m_{2}}\right)^{2}
\frac{V_{\|}}{4r^{2}} + \left(\frac{1}{m_{1}} + \frac{1}{m_{2}}\right)^{2}
\frac{V_{L}}{4r^{2}}\right]\bf (r \cdot p)( r \cdot p) .	
\end{equation}
\end{mathletters}
It is important to note that Eq.\ (\ref{com}) is a classical expression for 
the velocity-dependent potential energy of two classical sources, and thus it   carries no instructions for ordering the momentum and position operators.
Certainly, the quantum mechanical generalization of Eq.\ (\ref{com}) must
respect the principle of Hermiticity. Although this principle is an adequate
guide to determine the appropriate combination of any two noncommuting
operators, some additional considerations are required for more complicated
combinations. To resolve the ambiguity, we choose to interprete the combinations
of momentum and position operators in Eq.\ (\ref{com}) with the Gromes 
\cite{gr77} double-bracket notation,
\begin{equation}
\label{gromes}
\{\{p_{i}F({\bf r})p_{j}\}\} = \frac{1}{4} \left[ p_{i}Fp_{j} + p_{j}Fp_{i}
+ p_{i}p_{j}F + F p_{i}p_{j} \right].
\end{equation}
This expression has the virtue that it is manifestly Hermitian and that it
is involved in the nonrelativistic reductions of several covariant Dirac
potentials.

One additional difficulty requires some comment. The S-state expectation
value of $V_{v^{2}}$, the first-order perturbation theory improvement, 
leads to singular integrals. Such terms arise because of the dominant 
influence of the Coulomb part of $V_{0}$ upon the Salpeter wave function
at short distances. Nickisch, Durand
and Durand \cite{ndd84} have shown that the small r behavior of the 1S radial
function is given by $R_{1S}(r) \rightarrow B r^{-\rho }$, where $\rho \approx
0.25 $ for realistic values of $\alpha_{S} $. To illustrate this behavior we    
choose one of the factors that arises from the last term of $V_{rad}$, that is
\begin{equation}
\label{integ}
I_{L} = \int_{0}^{\infty } dr\,r^{2} \left(\frac{\partial R_{n\ell }}
{\partial r}\right)^{2} V_{L}. 
\end{equation}
The small $r$ behavior of the integrand is given by $ G(r) \rightarrow
-4 \alpha_{S} B^{2} \rho^{2} / 3 r^{1 + 2\rho}$, which leads to a divergent
integral at the lower limit. An important part of the reason for
this divergence can be traced to the nonrelativistic limit used
for the velocity operators. The relativistic expressions for the velocity
operators of Eq.\ (\ref{angrad}) are $v_{i} = p / \sqrt{m^{2}_{i} + p^{2}} $,
which have a finite large-momentum limit, in contrast to $v_{i}\approx p/m_{i}$,
which we used to obtain Eq.\ (\ref{com}). Thus, the constituent masses should
provide a high-momentum cutoff in a more accurate treatment. Translating this
observation into position space suggests that the Compton wavelengths of the
constituent quarks should provide a short-distance cutoff, which would
prevent the singular behavior. Since the singular behavior
arises in the perturbative limit of Eq.\ (\ref{dar}) and this equation
is symmetric in the two masses, we choose a cutoff that is the geometric mean
of the constituent Compton wavelengths, that is, $\Lambda = \left(m_{1}
m_{2}\right)^{1/2}$. This choice does not introduce any new paramters into the
calculation. Since all of the singular integrals that arise in the calculation
of the expectation value of $V_{v^{2}}$ have exactly the same behavior, one
cutoff suffices to eliminate this behavior.

\section{Results and Discussion}
\label{sec:res}
Our results for the 1S B meson masses are shown in Fig.\ \ref{fig1}, where they
are compared with the spin averages of the measured B meson masses\cite{pdg96}.
The full Salpeter curve shows the results of a calculation with the central
potential of Eq.\ (\ref{centpart}) only. This curve exhibits a uniform decrease,
as the light quark masses decreases, in contrast to the pathological behavior
\cite{lsg,mkn93} of the Schr\"{o}dinger equation results. It suggests that the
Salpeter equation is a reasonable starting point for a potential model 
calculation since it is not too far above the data. The remaining four curves of
Fig.\ \ref{fig1} include the effects of the velocity-dependent potential
$V_{v^{2}}$ of Eq.\ (\ref{angrad}). Each of these curves include perturbative
and nonperturbative effects since they use the same expressions for $V_{+}$, 
$V_{-}$ and $V_{\|}$. The dotted curve is based on the last of Eqs. (
\ref{rempot}) for $V_{L}$. Thus it includes all of the physics described by the
BBZ model. In contrast, the dashed curve includes only perturbative effects
in $V_{L}$, since it is based on Eq. (\ref{secondform}).
The relativistic correction of the dotted curve has the wrong sign to move the
Salpeter results towards agreement with experiment.
The appearance of a pronounced minimum near 0.325 GeV      shows the
return of nonphysical behavior similar to the Schr\"{o}dinger results, that 
is, the mass of the system increases as the constituent mass decreases. Such
behavior has also been noted by BBZ in an Appendix of Ref.\cite{ba95}. 
However, the results of the dashed curve exhibit the characteristics
necessary for agreement with experiment. The
dashed curve intersects the measured values near $ m_{u} = 0.180 \: {\rm GeV}$
and $m_{s} = 0.220 \: {\rm GeV}$, which are reasonable values         
for the constituent masses \cite{ballcom,sr80,gi85}.

The remaining two curves in Fig.\ \ref{fig1} are based on additional 
modifications of $V_{L}$ or $V_{v^{2}}$, which might be helpful in isolating
the reasons behind the striking differences between the dotted and dashed
curves there. The dot-short dashed curve of Fig.\ \ref{fig1} is based on
Eq. (\ref{thirdform}). All of the difference between this curve and the
dashed curve is due to the exponential damping of the Yukawa factor.
Such a pronounced dependence is a consequence of the $r^{2}$
weighting of the integrand of the cutoff form of the integral of
Eq. (\ref{integ}).

The long and short dashed curve is the result of an attempt to see if the
freedom implicit in mass renormalization can be helpful in removing the 
discrepancy between the BBZ model (dotted curve) and experiment. 
Barchielli, Brambilla and Prosperi \cite{bbp} show that a redefinition of
the constituent masses leads to changes in the components of the 
velocity-dependent potential $V_{v^{2}}$---at least in the nonrelativistic
limit---, since, if one makes the replacement $m_{i} \rightarrow m_{i} +
\Delta $, the Hamiltonian

\begin{equation}
\label{massr1}
H = m_{1} + m_{2} + \frac{p^{2}}{2 m_{1}} + \frac{p^{2}}{2 m_{2}}
  + V_{0} + V_{SD} + V_{v^{2}},
\end{equation}

is changed to

\begin{equation}
\label{massr2}
H_{R} = m_{1} + m_{2} + 2 \Delta + \frac{p^{2}}{2 m_{1}} + 
      \frac{p^{2}}{2 m_{2}} - \frac{\Delta }{2} \left( \frac{p^{2}}{m_{1}^{2}}
      + \frac{p^{2}}{m_{2}^{2}} \right) + V_{0} + V_{SD} + V_{v^{2}} .
\end{equation}

The two $\Delta$-dependent terms can be absorbed in a redefinition of the
potentials $V_{0}$ and $V_{v^{2}}$. However, one must be careful to respect
the momentum dependence prescribed by the form of Eq. (\ref{massr2}). This
requires that the same constant be subtracted from each of the component
potentials, $V_{+}$, $V_{-}$, $V_{\|}$ and $V_{L}$. The potential parameters
obtained by requiring the ``mass--renormalized '' component potentials to fit
the upsilon energy differences are identical to those obtained with the full
BBZ potentials, as expected. The differences between the long and short dashed
curve and the dotted curve,                                                     which become larger at smaller values of the light quark mass, probably
reflect the fact that one requires a different mass renormalization procedure
in this region.

It is worth noting that the parameters used to obtain each of the curves
in Fig.\ \ref{fig1} are slightly different, since in each case they
were obtained by fitting the energy levels of the heavy quark
systems. For example the dotted curve is based on the set,                      
$A = 0.229 \: {\rm GeV^{2}}$, $\alpha_{S} = 0.307 $, $m_{b} = 4.717\: {\rm GeV}$and $m_{c} = 1.308 \: {\rm GeV}$,
which differ slightly from the values listed in Eq.\ (\ref{paramet}), the set
used to calculate the dashed curve. The potential parameters for the 
two remaining curves in Fig.\ \ref{fig1} are identical to those for the
dotted curve.

The results for the 1S D meson masses depicted in Fig.\ \ref{fig2} are very
similar to those of Fig.\ \ref{fig1}, since each curve there has the same
qualitative features as its counterpart in Fig.\ \ref{fig1}. The dashed
curve in Fig.\ \ref{fig2} intersects the measured masses near $m_{u} = 0.130
\: GeV$ and $m_{s} = 0.200 \: GeV$. These values differ somewhat form those
determined in Fig.\ \ref{fig1}, but such differences could easily be accounted
for by a running mass. 

Results for the P state masses of the D mesons are presented in 
Fig.\ \ref{fig3}.
The dotted curve has a shallow minimum around 0.275 GeV, in qualitative
accord with the S state behavior. The dashed curve again intersects the data
points, but the values that one obtains for the constituent masses, namely
$m_{u} = 0.110 \; GeV$ and $m_{s} = 0.125 \; GeV$ are uncomfortably close 
together. The remaining two dashed curves in Fig.\ \ref{fig3} have 
qualitative differences from their S state counterparts.

Thus, our conclusion is that we have not been able to use the full BBZ
potentials to simultaneously account for the energies of the heavy quark
systems and the heavy flavour systems with a single set of potential
parameters in the context of the spinless Salpeter equation. It will be very
interesting to see if calculations based upon the formalism of Hardekopf
and Sucher \cite{su84} will remove this discrepancy.	

Since lattice gauge calculations are becoming more and more accurate, it is
worthwhile to examine the recent calculations \cite{bal97}      
of Bali, Schilling and Wachter
(BSW) to see if they contain any guidance about which form of $V_{L}$ is best to
use. BSW present results for the velocity-dependent spin-independent corrections
$V_{b}$, $V_{c}$, $V_{d}$, $V_{e}$, which are defined as in the recent work of
Brambilla and Vairo \cite{br97}. These component potentials are 
certain linear combinations of
the potentials $V_{+}$, $V_{-}$, $V_{\|}$, $V_{L}$. For example, $V_{c}$ and
$V_{e}$ are defined by the relations,

\begin{mathletters}
\begin{equation}
V_{c} = \left( V_{-} + V_{L} - V_{+} - V_{\|} \right) / 2,
\end{equation}

\begin{equation}
V_{e} = \left( V_{+} + V_{-} - V_{\|} - V_{L} \right) / 4.
\end{equation}
\end{mathletters}

In Fig.\ \ref{fig4} we compare two of our determinations of $V_{c}$ with 
representative lattice data taken from Fig. 15 of Ref.\ \cite{bal97}. The 
dotted curve there is based on the use of Eq.\ (\ref{rempot}) for $V_{L}$ and
the paramter set listed above in this section. The dashed curve there
is based on the use of Eq.\ (\ref{secondform}) for $V_{L}$ and the parameter
set of Eq.\ (\ref{paramet}). 
The lattice data does seem to favor the dotted curve over the 
dashed curve, but the error bars in the lattice data are large enough to
prevent any strong indication. It is worth noting how much more sensitive the
energy level calculations are to the differences between these two cases than
the potential of Fig.\ \ref{fig4}.

\newpage
\begin{figure}
\caption{Masses for 1S B mesons as a function of the light quark constituent 
mass.}
\label{fig1}
\end{figure}

\begin{figure}
\caption{Masses for 1S D mesons as a function of the light quark
constituent mass.}
\label{fig2}
\end{figure}
 
\begin{figure}
\caption{Masses for 1P D mesons as a function of the light quark
constituent mass.}
\label{fig3}
\end{figure}

\begin{figure}
\caption{Comparison of the lattice results for $V_{c}$ with those based on
two different forms of $V_{L}$ used in our calculations.}
\label{fig4}
\end{figure}

\widetext   
\begin{table}
\caption{Matrix elements of the residual potential function $f(r)$ and S-state  energy eigenvalues for the upsilon system. The parameters used for
this calculation are those listed in Eq.\ (\ref{paramet}) and $\beta = 2.0 \:
{\rm GeV}$.}
\label{potmat}
\begin{tabular}{ccccccc}
Matrix elements& $5 \times 5$& $10 \times 10$& $20 \times 20$& $30 \times 30$
& $40 \times 40$\tablenote{The Hamiltonian matrix used for the eigenvalues in   the last two columns was $20 \times 20$.} & Analytic results\\
\tableline
$(f)_{00}$& 0.38648& 0.38648& 0.38648& 0.38648& 0.38648& 0.38648\\
$(f)_{01}$& 0.03346& 0.03346& 0.03346& 0.03346& 0.03346& 0.03346\\
$(f)_{02}$& 0.00318& 0.00318& 0.00318& 0.00318& 0.00318& 0.00318\\
$(f)_{03}$& 0.00031& 0.00031& 0.00031& 0.00031& 0.00031&        \\
$(f)_{11}$& 0.35304& 0.35304& 0.35304& 0.35304& 0.35304& 0.35304\\
$(f)_{12}$& 0.04799& 0.04799& 0.04799& 0.04799& 0.04799& 0.04799\\
$(f)_{33}$& 0.30244& 0.30264& 0.30264& 0.30264& 0.30264&        \\
$(f)_{44}$& 0.26930& 0.28336& 0.28336& 0.28336& 0.28336&        \\
$E(1S)$& 9.47430& 9.47368& 9.47344& 9.47340& 9.47341& \\
$E(2S)$& 10.00877& 10.00718& 10.00703& 10.00704& 10.00701& \\
$E(3S)$& 10.38178& 10.36697& 10.36683& 10.36683& 10.36682& \\
\end{tabular}
\end{table}

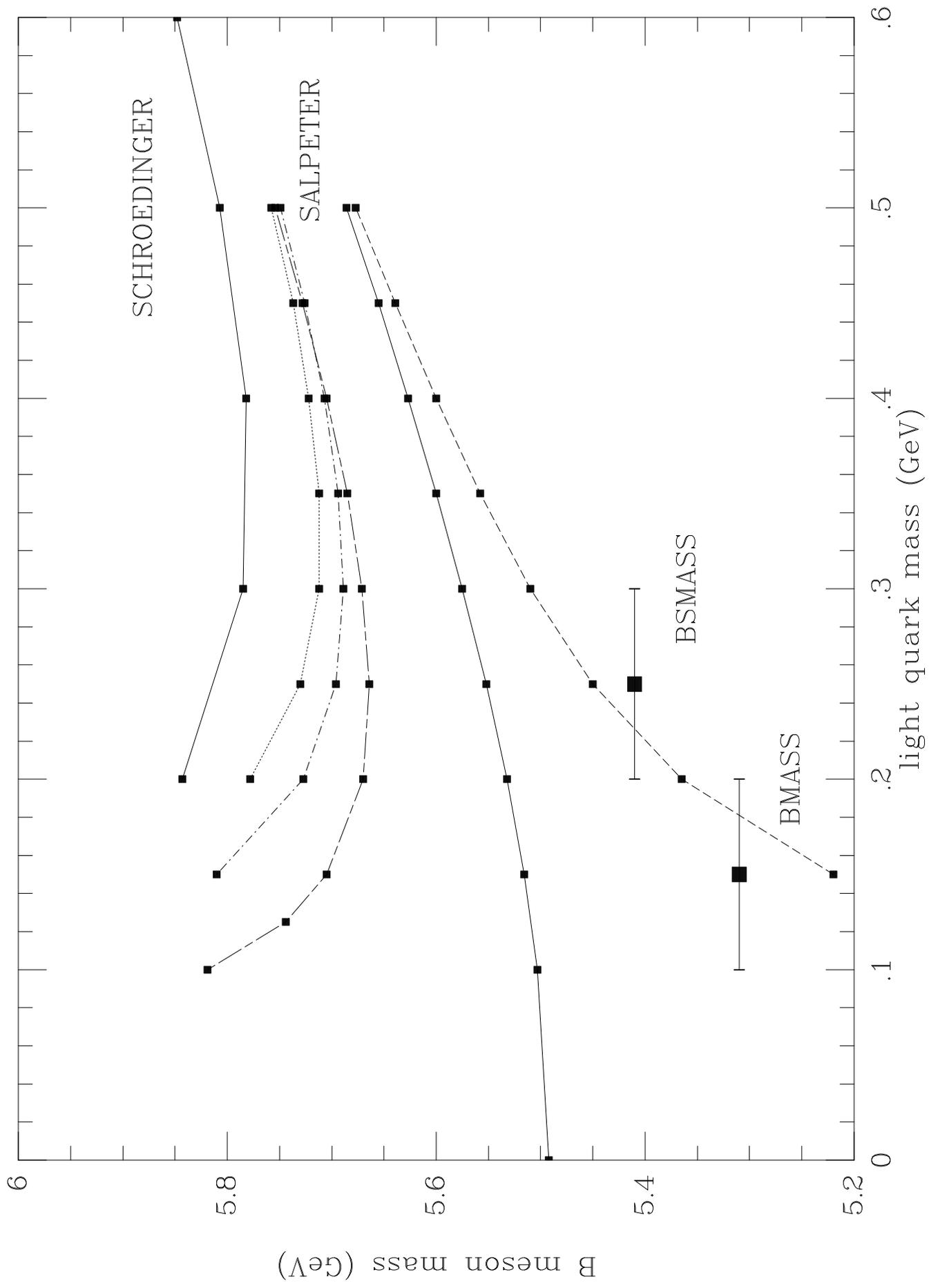
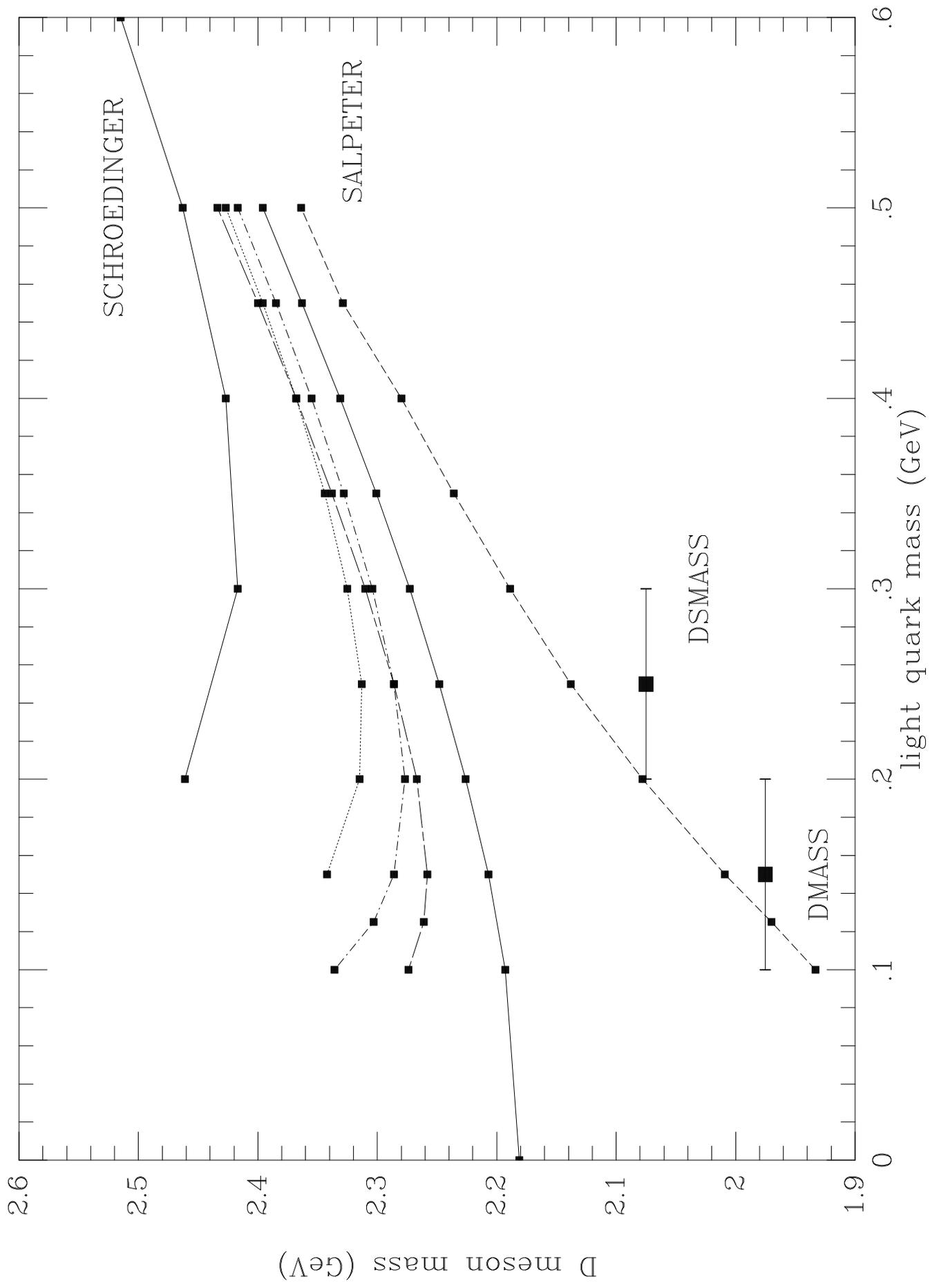
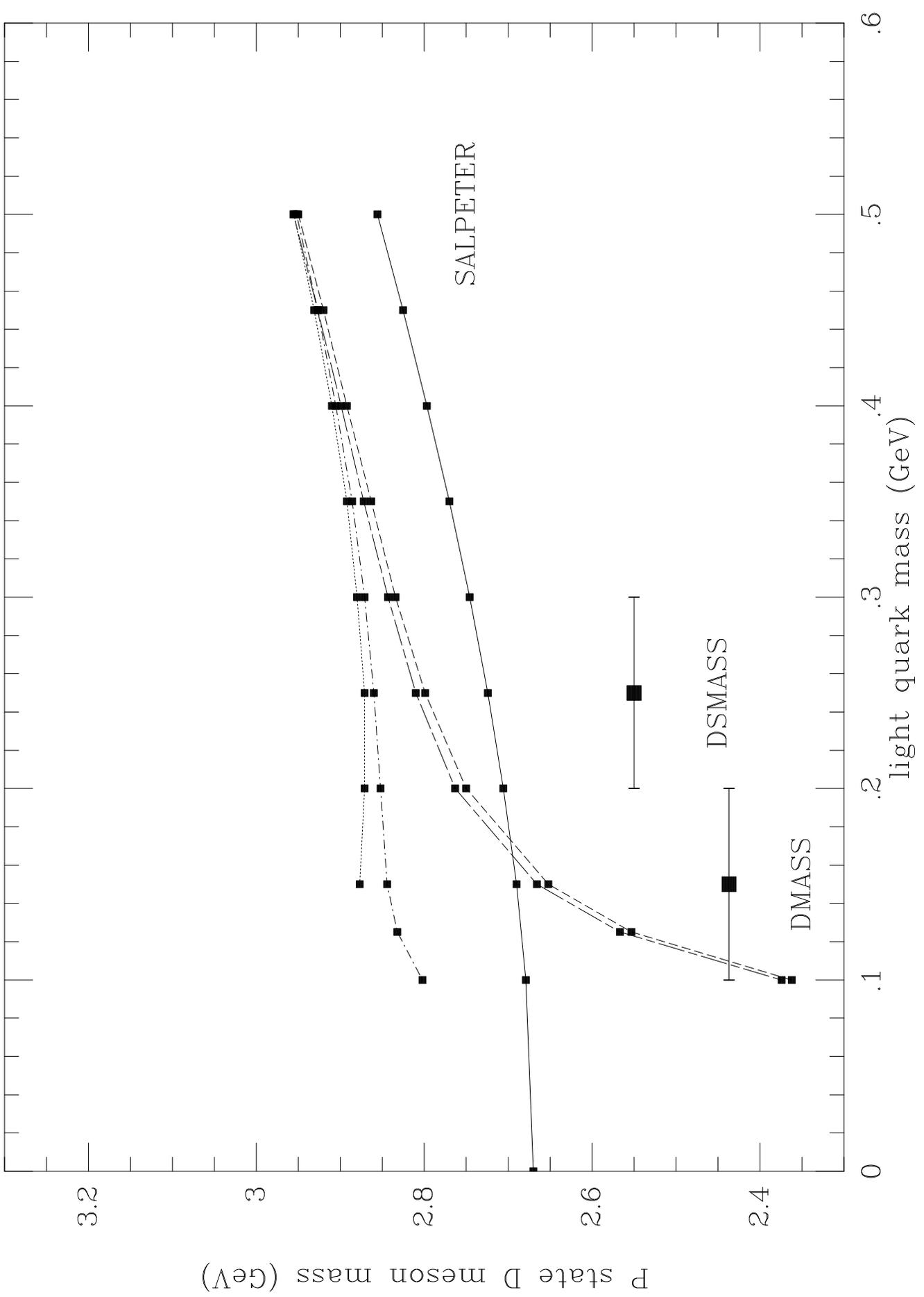
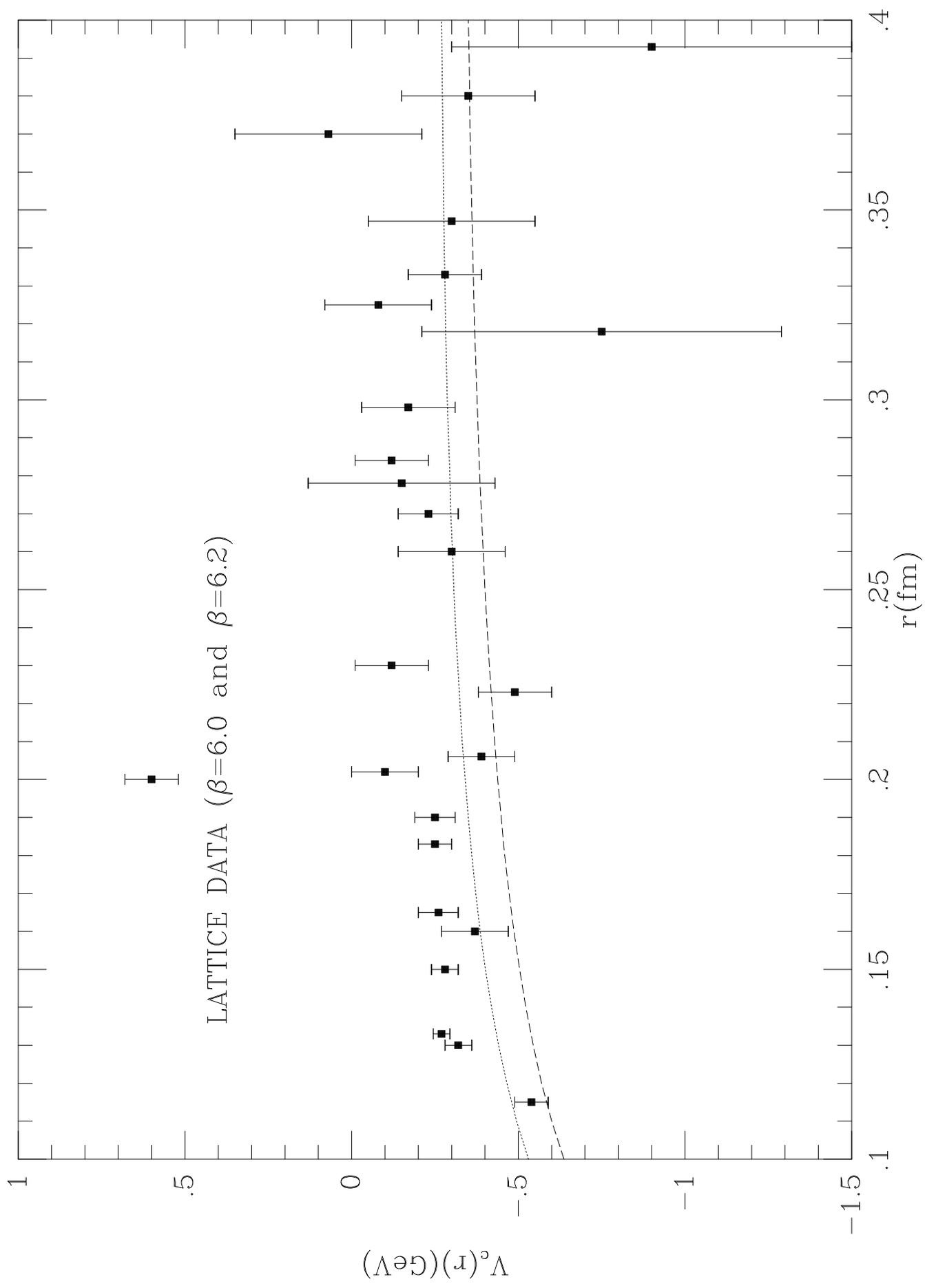
\end{document}